\documentclass[letterpaper, 10 pt, conference]{ieeeconf}
\IEEEoverridecommandlockouts
\usepackage{cite}
\usepackage{amsmath,amssymb,amsfonts}
\usepackage{algorithmic}
\usepackage{algorithm} 
\usepackage{algorithmic}
\usepackage{graphicx}
\usepackage{textcomp}
\usepackage{xcolor}
\usepackage{amsmath}
\usepackage[textsize=footnotesize]{todonotes}
\newtheorem{theorem}{Theorem}[section]

\newtheorem{lemma}[theorem]{Lemma}
\newtheorem{example}{\indent Example}
\def\and{\textrm{ and}}

\newcommand{\PWA}{\mathrm{PWA}}

\newcommand{\genpwaparamM}{A}
\newcommand{\genpwaparamV}{a}
\newcommand{\pMatrix}{E} % general partition matric
\newcommand{\pVec}{e} % general partition vector
\newcommand{\Mode}{I}
\newcommand{\state}{x}

\newcommand{\pState}{X}

\newcommand{\Vertex}{\mathcal{F}_0}
\newcommand{\Edge}{\mathcal{F}_1}
\newcommand{\SignFunction}{sgn_{\dot V}}

\newcommand{\prtition}{\mathcal{P}}
\newcommand{\VectorField}{\dot x}
\newcommand{\derivative}{\dot V}

\newcommand{\qState}{Z}

\newcommand{\vMatrix}{E}
\newcommand{\vVec}{e}

\newcommand{\Rprtition}{\mathcal R}

\newcommand{\Pindex}{\Mode(\prtition)}

\newcommand{\Int}[1]{\mathrm{Int}\left(#1 \right)}
\newcommand{\Dom}[1]{\mathrm{Dom}\left(#1 \right)}
\newcommand{\conv}[1]{\mathrm{conv} \left(#1 \right)}

\newcommand{\R}{\mathbb{R}}

\usepackage[normalem]{ulem} % use normalem to protect \emph
\usepackage{hyperref}
\hypersetup{
colorlinks=true,
linkbordercolor=orange,
citebordercolor=blue,
citecolor=blue,
filecolor=magenta,
urlcolor=magenta,
urlbordercolor=blue
}
\DeclareMathOperator*{\argmax}{argmax}
\title{\LARGE \bf Automated Stability Analysis of Piecewise Affine Dynamics Using Vertices}

\author{Pouya Samanipour and Hasan A. Poonawala
\thanks{Pouya Samanipour and Hasan A. Poonawala are with the Department of Mechanical and Aerospace Engineering, University of Kentucky, Lexington, USA
{\tt\small\{samanipour.pouya,hasan.poonawala\}@uky.edu}. The corresponding author is Hasan A.Poonawala. 
This work is supported by the Department of Mechanical Engineering at the University of Kentucky.}}
\begin{document}
\maketitle
\begin{abstract}
This paper presents an automated algorithm to analyze the stability of piecewise affine ($\PWA$) dynamical systems due to their broad applications. 
We parametrize the Lyapunov function as a $\PWA$ function, with polytopic regions defined by the $\PWA$ dynamics. Using this parametrization, Stability conditions can be expressed as linear constraints restricted to polytopes so that the search for a Lyapunov function involves solving a linear program. However, a valid Lyapunov function might not be found given these polytopic regions.
A natural response is to increase the size of the parametrization of the Lyapunov function by dividing regions and solving the new linear program. %
% To overcome this challenge, we present a new algorithm that allows the size of the parametrization of the Lyapunov function to increase by dividing regions during the search process. 
This paper proposes two new methods to divide each polytope into smaller ones. 
The first approach divides a polytope based on the sign of the derivative of the candidate Lyapunov function, while the second divides it based on the change in the vector field of the $\PWA$ dynamical system. 
In addition, we propose using Delaunay triangulation to achieve automated division of regions and preserve the continuity of the $\PWA$ Lyapunov function. 
Examples involving learned models and explicit MPC controllers demonstrate that the proposed method of dividing regions leads to valid Lyapunov functions with fewer regions than existing methods, reducing the computational time taken for stability analysis. %
\end{abstract}
% \begin{IEEEkeywords}
% Single-hidden layer ReLU, Stability analysis, Piecewise affine dynamics, Lyapunov Function, Optimization
% \end{IEEEkeywords}
\section{Introduction}
% \subsection{Motivation}
Piecewise affine ($\PWA$) dynamical systems have gained popularity in robotics \cite{marcucci2017approximate} and the automotive industry \cite{sun2019hybrid} due to their wide applications. $
\PWA$ concepts are utilized in advanced controllers, including gain-scheduled flight control systems\cite{bemporad2000piecewise} and Takagi-Sugeno fuzzy systems\cite{qiu2020fuzzy}. Affine systems with control saturation can be expressed using $\PWA$ dynamics, enabling effective synthesis of controllers through explicit model predictive control (MPC)\cite{alessio2009survey}. However, obtaining a Lyapunov function for stability guarantees with explicit MPC can be challenging. Alternatively, there is an increasing trend in using supervised machine learning methods for learning dynamics and controllers\cite{elguea2023review,jebellat2021training}. Neural networks (NN) with the rectified linear unit (ReLU) activation functions have been employed to convert closed-loop dynamics into PWA dynamics\cite{samanipour2023stability}. The stability of these methods, however, is not guaranteed, emphasizing the need to develop an automated approach to finding Lyapunov functions for learned models, including ReLU networks and explicit MPC.

Sampling-based methods \cite{abate2020formal,korda2022stability,ravanbakhsh2017learning} are prevalent for learning Lyapunov functions.  The Lyapunov function is learned from finite samples, and this function must meet the stability conditions at all states, therefore verification is a critical component of the analysis.
Verification can be performed in an inexact manner using relaxed convex problems \cite{fazlyab2020safety} or in an exact manner using Satisfiability Modulo theories (SMT) and Mixed-Integer Programs (MIP)\cite{abate2020formal,chen2021learning,dai2021lyapunov,zhou2022neural}. The exact verifier certifies the Lyapunov function or generates counterexamples violating the stability conditions. Counterexamples can be incorporated into training samples for iterative learning. However, the computational complexity of the verifier remains a challenge.

An alternative to the learning approach is to parameterize the Lyapunov function and solve it as an optimization problem \cite{anderson2015advances,iervolino2017lyapunov,iervolino2020asymptotic}. The Sum of Squares (SOS) method is employed to find the Lyapunov function for nonlinear dynamics \cite{anderson2015advances}, but it can be computationally complex. A piecewise quadratic (PWQ) parameterization of the candidate Lyapunov function is proposed in\cite{iervolino2020asymptotic,iervolino2017lyapunov}. However, these methods must deal with the conservatism of the S-procedure, and the results are limited to two-dimensional examples.
%Alternative to the learning approach is parameterizing the Lyapunov function and searching for the solution by forming an optimization problem \cite{iervolino2017lyapunov,iervolino2020asymptotic,johansson1999piecewise,anderson2015advances}. The Sum of Square(SOS) is used to find the Lyapunov function for nonlinear dynamics in \cite{anderson2015advances}. The main challenge in SOS, however, is the complexity of computation. Parameterizing the candidate Lyapunov function as a piecewise quadratic(PWQ) is proposed in \cite{iervolino2017lyapunov,iervolino2020asymptotic}. 
%Utilizing the piecewise quadratic function requires dealing with linear matrix inequalities as well as S-procedure conservatism.
%It was demonstrated in \cite{iervolino2017lyapunov} that stability conditions can be less conservative by using cone-constrained inequalities to define stability conditions.  Moreover, a Filippov solution that includes single-mode Caratheodory, sliding mode, and forward Zeno behavior is proposed in \cite{iervolino2020asymptotic}. %There is no requirement that the proposed discontinuous Lyapunov function takes the form of a PWQ function. 
% A less conservative approach was proposed in \cite{iervolino2017lyapunov}, but the results were limited to two-dimensional examples.
%PWA

Instead of relying on the PWQ Lyapunov function, \cite{johansson1999piecewise} utilized the $\PWA$ function to parameterize the Lyapunov function. An algorithm has been developed for finding a $\PWA$ Lyapunov function using partition refinement in \cite{9030067}. A method for calculating the Lyapunov function for conewise $\PWA$ dynamics was proposed by \cite{poonawala2021stability}. The dynamics and controller of $\PWA$ have been parameterized as a ReLU in \cite{samanipour2023stability}. The Lyapunov function and the controller are found by parameterizing Lyapunov conditions as quantifier-free constraints for a bilinear quadratic optimization problem\cite{samanipour2023stability}. Although the Lyapunov condition for the $\PWA$ Lyapunov function can be expressed without conservatism, the PWQ Lyapunov function receives more attention in the literature. 

The refinement process in the context of Lyapunov stability analysis presents several challenges, such as preserving the continuity of the candidate Lyapunov function and dividing complex polytopes effectively. 
We propose the following contributions to address the challenges in the refinement and continuity of the $\PWA$ Lyapunov function.
% Here, we propose new refinement techniques in order to implement refinement more effectively. We also use a tool to automate the process of refinement and address the continuity issue. 
\paragraph{Contributions}
The paper introduces two novel methods for dividing cells during the search for valid Lyapunov functions. The first method utilizes the derivative of the Lyapunov function as a criterion to divide a cell, while the second method analyzes the vector field of the $\PWA$ dynamics to do so. By examining the behavior of the Lyapunov function derivative or vector field, these methods determine suitable locations for proposing new vertices that will define new cells, since we use the vertex representation for polytopes. Furthermore, the paper proposes using Delaunay triangulation to automate the refinement process for cells. The proposed refinement methods offer the advantage of finding valid Lyapunov functions with fewer refinements compared to existing techniques. The efficacy of the search procedure is demonstrated through non-trivial examples, where valid Lyapunov functions are successfully identified within reasonable computation times. Additionally, the paper evaluates the effectiveness of the proposed approach in determining the region of attraction (ROA) by comparing the results with other methods. The comparison showcases the capability of the proposed approach to identify the ROA using the $\PWA$ Lyapunov functions. The contributions of this paper improve the refinement process addressing challenges in the parameterization of the Lyapunov function. %

% Following is a description of how this paper is organized. To begin with, a brief overview of the $\PWA$ and Lyapunov function parameterization is presented in Section~\ref{sec:Prelim}. In Section~\ref{sec:Main}, the optimization problem to find the Lyapunov function is established, and the refinement process is presented. The results, comparisons, and limitations are provided in Section~\ref{sec:res}. As a final step, the conclusion is presented. 
% We make the following contributions:
% \begin{enumerate}
%     \item The main contribution of this paper is using Delaunay triangulation to refine the partitions. This technique allows the refinement process to generalize to dimensions higher than two or three.  
%     \item 
%   \item In order to add more flexibility to the search for the valid Lyapunov function, We utilized the Delaunay triangulation to refine each partition with one of its neighbors simultaneously. 
%     \item To ensure that refinement will not lead to mixing dynamics, the midpoint was given to Delaunay triangulation. 
% \end{enumerate}
% A detailed description is provided in \ref{subsec:Delaunay}.
\section{Preliminaries} \label{sec:Prelim}
In this paper, we examine the stability analysis problem for dynamical systems described by piecewise affine functions as follows:
\begin{align}
  \dot x = \PWA(x).\label{eq:dynamicalsystem}
\end{align} 
% \end{align} 
where $x \in \mathbb{R}^n$ is the state variable, and term $\PWA$ denotes a piecewise affine function. We focus on continuous $\PWA$ functions with polytopic cells. 
% 
% In this paper, $\PWA$ Lyapunov functions are proposed as a method for analyzing stability properties of the closed-loop $\PWA$ dynamics \eqref{eq:dynamicalsystem}. The advantage of this approach is that it enables the use of linear optimization techniques for stability analysis. In addition, this method can also be used to examine the stability of ReLU NNs, which can be represented as $\PWA$ functions\cite{samanipour2023stability}. 
%  The aim of this paper is to consider the \textbf{stability analysis} problem for dynamical systems that have the following form: 
% \begin{align}
%   \dot x = \PWA(x),\label{eq:dynamicalsystem}
% \end{align} 
% where $x \in \mathbb{R}^n$ is the state variables, and term $\PWA$  denotes a piecewise affine function. %
% Section~\ref{sec:prelims} defines these functions more precisely. %
% For now, we mention that we focus on continuous $\PWA$ functions with polytopic cells. % 
% We will use candidate Lyapunov functions $V(x)$ that are parametrized as $\PWA$ functions. %
% Parameterizing the candidate Lyapunov function as a $\PWA$ will help us to deal with closed-loop properties such as stability, asymptotic stability, and exponential stability using Linear optimization. %
% The emphasis is, however, placed upon asymptotic stability in order to simplify the explanation. %
% Also, this method may be used to analyze the stability of ReLU NNs since $\PWA$ functions can be derived from this type of network, as described in \cite{samanipour2023stability}.  
% \section{Representations of Piecewise Affine Functions}
\label{sec:prelims}
The rest of this section formally describes $\PWA$ functions.
%We refer to the standard representation of $\PWA$ functions (see Section \ref{sec:pwafun}) as an \textbf{explicit parametrization}\cite{samanipour2023stability}.
\paragraph*{\bf Notation}
An index for each element in the set $S$ constitutes the set $\Mode(S)$.
%$\mathbf{x}_S$ denotes a set of variables $\{x_i \}_{\mode \in \Mode(S)}$, while $\mathbf{x}_I$ denotes a set of variables $\{x_i \}_{\mode \in \Mode}$. 
The convex hull, the interior, the boundary, and the closure of the set $S$ are denoted by $\conv{S}$, $\Int{S}$, $\partial S$, and $\overline{S}$ respectively.
The transpose of matrix $A$ is $A^T$. $\langle \cdot,\cdot\rangle$ denotes the inner product, $\angle(\cdot,\cdot)$ is the angle between two vectors, and $|\cdot|_2$ is the standard $L_2$ norm. 
It should be noted that the symbol $\succeq$ is the element-wise version of $\geq$.
% \paragraph*{\bf Definition} 
\subsection{Partitions And Refinements}
%A partition $\prtition$ is a collection of subsets $\{\pState_i \}_{i \in \Pindex}$, \mcr{where $\pState_i \subseteq \R^n$, $n \in \mathbb{N}$, and $\overline{\Int{\pState_i}} =\pState_i$ for each $i \in \Pindex$. %, Int(\pState_i) \neq \emptyset$ 
%Furthermore,} $\Int{\pState_i} \cap  \Int{\pState_j} = \emptyset$ for each pair $i,j \in \Pindex$ such that $i \neq j$. %
%We refer to $\cup_{i \in \Pindex} \pState_i$ as the domain of $\prtition$, which we also denote by $\mathrm{Dom}(\prtition)$. %
%We also refer to the subsets $\pState_i$ in $\prtition$ as the cells of the partition. %
%We assume that there exists a neighborhood of $x$ that intersects with only a finite number of cells in $\prtition$, for each $x \in \mathrm{Dom}(\mc P)$. %

%Let $\prtition = \{ Y_i\}_{i \in I}$ and $\Rprtition = \{ \qState_j\}_{j \in J}$ be two partitions of a set $S = \Dom{\prtition} = \Dom{\Rprtition}$. %$\State$. %
%A partition $\Rprtition$ is a \emph{refinement} of $\prtition$ if $\qState_j \cap Y_i \neq \emptyset$ implies that $\qState_j \subseteq Y_i$. %
%We denote the set of refinements of a partition $\prtition$ as $\mathrm{Ref}(\prtition)$. %
In this paper, we define a partition $\prtition$ as a collection of subsets $\{\pState_i \}_{i \in \Pindex}$, where each $\pState_i$ is a closed subset of $\R^n$ and $int(X_i)\cap int(X_j)=\emptyset$, $\forall i, j \in\Pindex$ and $i\neq j$. The domain of the partition, $\mathrm{Dom}(\prtition)$, is the union of all the cells in $\prtition$. 
%If $x$ is a point in $\mathrm{Dom}(\mc P)$, then a neighborhood intersects only a limited number of cells in $\prtition$ for any given point.

Given two partitions $\prtition = \{ Y_i\}_{i \in I}$ and $\Rprtition = \{ \qState_j\}_{j \in J}$ of a set $S = \Dom{\prtition} = \Dom{\Rprtition}$, we say that $\Rprtition$ is a refinement of $\prtition$ if $\qState_j \cap Y_i \neq \emptyset$ implies that $\qState_j \subseteq Y_i$. We denote the set of all refinements of $\prtition$ as $\mathrm{Ref}(\prtition)$\cite{samanipour2023stability}. 
\subsection{Piecewise Affine Functions}
\label{sec:pwafun}
We explicitly parameterize a piecewise affine function $\PWA(x)$ by a partition $\prtition = \{\pState_i \}_{i \in \Pindex}$ and a collection of matrices $\mathbf A_\prtition = \{\genpwaparamM_i \}_{i \in \Pindex}$ and vectors  $\mathbf a_\prtition = \{\genpwaparamV_i \}_{i \in \Pindex}$ such that 
\begin{align}
\PWA(x)	&= \genpwaparamM_{i} \state + \genpwaparamV_{i}, \textrm{ if } \state \in \pState_i,  \text{ where}\nonumber\\
  \pState_i &= \{x \in \R^n \colon \vMatrix_i \state + \vVec_i  \succeq 0 \} \label{eq:definePWAfun}.%, \text{ or}\label{eq:lyapsets}\\
\end{align}
Note that a generic $\PWA$ function may not be continuous unless we appropriately constrain the parameters $\genpwaparamM_i$, $\genpwaparamV_i$, $\pMatrix_i$, and $\pVec_i$\cite{poonawala2021training,johansson1999piecewise}. % 
It is assumed that any $\PWA$ function in this paper with this explicit form meets such constraints and is always continuous. Additionally, we consider the origin to be the equilibrium, thus denoting index sets $I_0$ and $I_1$ for cells containing and not containing the origin respectively. Also, we assume that all the cells are bounded. Therefore, we can use the vertex representation for all cells. A vertex is a facet of dimension $0$ for a cell~\cite{henk2017basic}. Each cell of a partition can be represented using its vertices:
\begin{equation}
     X_i= {\conv{\Vertex(X_i)}},
\end{equation}
where $\Vertex(X_i)$ represents the set of vertices of the cell $X_i$.   
\section{Main Algorithm}   
This section presents an overview of the stability analysis algorithm, which aims to construct an optimization problem to discover the $\PWA$ Lyapunov function. The algorithm consists of two main components: the formulation of an optimization problem to find a valid Lyapunov function and a refinement process to enhance the flexibility of the $\PWA$ Lyapunov function. A detailed description of these components is provided in the subsequent section. For better comprehension, a pseudo-code representation of the algorithm is presented in \textbf{Algorithm} \ref{alg:refinesearch}. The termination condition of the algorithm is determined by two criteria: either a valid Lyapunov function is found, or the optimization process exceeds the predefined timeout threshold of 3600 seconds. It should be emphasized that in the case of unstable systems, the algorithm needs to be manually terminated.
\begin{algorithm}[tb]
    \begin{algorithmic}
    \REQUIRE  $\PWA(x)$, Vertices
    \STATE{Solve Optimization Problem \eqref{eq:lyaprelaxfinal} with the initail $\PWA$ dynamics.}
    \WHILE{$\Sigma_{i=1}^{N}\tau_i\neq 0$ or Computational time $\leq3600$(sec)}  
    \FOR{$i \in I_s$} \STATE{1- Finding new vertices using presented methods(see \ref{sec:new vertices}).}
    \STATE{2- Add the new vertex to $B$ \eqref{eq:Buffer}.  }
    \ENDFOR
    \FOR{$i \in I_{split}$}
     \STATE{1- Finding set of vertices $\mathcal{V}_{new}(i)$ for cell $X_i$.}
     \STATE{2- Forming sub-cells $DT(\Vertex(X_i) \cup \mathcal{V}_{new}(i))$ (see (\ref{sec:Delauney})).}
     \ENDFOR
%	\STATE{$\lyapH \gets \bmat{\lyapH \\ H^{new}\\ -H^{new}} $, $\lyapb \gets \bmat{\lyapb \\ b^{new}\\ -b^{new}} $}
    \STATE{Solve Optimization Problem \eqref{eq:lyaprelaxfinal} with the refined $\PWA$ dynamics.}
     \ENDWHILE
%	\STATE{$\vprtition \gets \Rprtition_m$}
    \RETURN $V(x) = p_i^Tx+q_i    \quad \forall i\in I(\prtition) $.
    \end{algorithmic}
    \vspace{1em}
    \caption{Verifying Stability using Vertices}
    \label{alg:refinesearch}
\end{algorithm} 
\section{Optimization based search for Lyapunov function}\label{sec:Main} 
In this section, first, we describe the general idea of the stability analysis and the Lyapunov function. In the next step, we parameterize the Lyapunov function as a $\PWA$ function. Then we present the stability condition for $\PWA$ dynamics with a $\PWA$ candidate Lyapunov function. In \ref{Relaxation}, We convert the stability analysis problem to a linear optimization problem. We construct the optimization problem to be always feasible; however, only a specific solution is accepted as a valid Lyapunov function. Furthermore, we proposed new refinement approaches \ref{subsec:Delaunay} to increase the capacity of the candidate Lyapunov functions, facilitating the search for valid Lyapunov functions. 
\subsection{Lyapunov function}
The Lyapunov stability theory is well known for its application to the analysis of nonlinear dynamical systems~\cite{khalil2015nonlinear}.
Assume that $V:D\rightarrow R$ is a continuously differentiable function, and $x=0$ is the equilibrium point of equation~(\ref{eq:dynamicalsystem}). In this case, equation~(\ref{eq:dynamicalsystem}) will be asymptotically stable if and only if $V$ is strictly positive definite and strictly decreasing $\forall x \in D-\{0\}$. 
\subsection{$\PWA$ Lyapunov function}
In the paper, we investigate the use of $\PWA$ Lyapunov functions on a bounded partition that aligns with the $\PWA$ dynamics \eqref{eq:definePWAfun} structure. This assumption can be used to further reduce computation costs by taking advantage of the convexity property. Specifically, if all cells in the partition are bounded, an affine function is considered positive on a particular cell $X_i$ if and only if it is positive on all vertices of $X_i$\cite{johansson1999piecewise}. This observation allows for simplified analysis and computation of the Lyapunov function. 
%Using this parameterization, we are able to formulate the following conclusion. 
Consider a candidate $\PWA$ Lyapunov function such that:
\begin{align}   \label{Eq:Lyap PWA}
    V(x)=\left\{
   \begin{array}{lr}
        p_i^Tx+q_i &\quad \text{for} \quad i \in I_1 \\
        p_i^Tx &\quad \text{for} \quad i \in I_0.
   \end{array}
   \right.
\end{align}
In the equation above, the function $V(x)$ is continuous and differentiable in the interior of the cell. It is possible to calculate the derivative of the candidate Lyapunov function, along the dynamic, $ \dot x=f(x) $, in the interior of cell $X-\{0\}$:
\begin{equation}
    \mathcal L_f V =  \langle \nabla V , f(x) \rangle ,\label{eq:generallyapcond}
\end{equation}
where $\nabla V$ is the gradient of $V(x)$, and $\mathcal L$ is the lie derivative.

When $V(x)$ is differentiable at $x$, let the local affine Lyapunov function be $V(x) = p^T x + q$, and the dynamics be $\dot x = A x + a$. The derivative of the Lyapunov function along the trajectories can be calculated as follows:
\begin{equation}\label{eq:derivative definition}
 \dot V=p^T(Ax+a).
 \end{equation}
\begin{lemma}
Let $\{X_i\}_{i\in I}$ be a partition of a bounded subset of $\mathbb{R}^n$ into convex polytopes with vertices $v_k$.  
\begin{enumerate}
    \item The Lyapunov function (\ref{Eq:Lyap PWA}) will be positive definite iff:
        \begin{align}   \label{Eq:PD}
            &p_i^T v_k+q_i>0 \hspace{0.2em} &\text{for} \quad i \in I_1, v_k \in \Vertex(X_i)\nonumber\\
            &p_i^T v_k>0 &\quad \text{for} \quad i \in I_0, v_k \in \Vertex(X_i).
        \end{align}
    \item $\dot{V}$, \eqref{eq:derivative definition}, will be  a negative definite function iff:
    \begin{align}   \label{Eq:ND}
        p_i^T  (A_i  v_k+a_i) <0 &\hspace{0.3em} \text{for} \hspace{0.3em} i \in \Pindex, v_k \in \Vertex(X_i).
    \end{align}
\end{enumerate}   
\end{lemma}
\begin{proof}
These results can be derived directly as a result of parameterizing the candidate Lyapunov function in the affine form. 
\end{proof}
The last step is to force the Lyapunov function to be continuous. To achieve this goal, the candidate Lyapunov function (\ref{Eq:Lyap PWA}) must meet the following requirements.
\begin{align}\label{eq:Equality}
    &V_i(v_k)=V_j(v_k)  ,\hspace{0.5em} i\neq j \in \Pindex, v_k \in \Vertex(X_i)\cap \Vertex(X_j).
\end{align}
\begin{theorem} \label{Th:General idea}
A Lyapunov function (\ref{Eq:Lyap PWA}) in the partition $\prtition$ is considered valid if there exist $p_i$ and $q_i$ satisfy  (\ref{Eq:PD})-(\ref{eq:Equality}) for every $v_k \neq 0 $. 
\end{theorem}
\begin{proof}
In this formulation \eqref{Eq:PD} guarantees that the Lyapunov function will be positive definite. Additionally, \eqref{eq:Equality} guarantees continuity. In this case, the Lyapunov function is Liptchitz continuous, but it is not differentiable at the boundary. As a result of the Lipchitz continuity of the Lyapunov function, we are able to use the Clarke generalized gradient and Clarke generalized derivative\cite{della2019smooth,baier2012linear}. According to Clarke generalized gradient $\partial V(x)$ for the $\PWA$ Lyapunov function \eqref{Eq:Lyap PWA} can be described as:
\begin{equation}
    \partial V(x)=conv(\{p_i:i\in I(p),x\in X_i\})
\end{equation}
The Clarke generalized derivative along $F$ for the differential inclusion $\dot x \in F(x)$ is provided by \cite{della2019smooth}
\begin{equation}\label{eq:Clarke derivative}
    \dot V_F=\{p^Tf:p\in \partial V(x),f\in F(x)\}.
\end{equation}
For points $x\neq0$ if $F$ is a singleton function then \eqref{Eq:ND} guarantees that $\dot V_F<0, \hspace{0.3em} \forall  p \in \partial V(x)$. As shown by \cite{baier2012linear}, the maximum of the \eqref{eq:Clarke derivative} upper-bounded the decrease of the Lyapunov function along solutions of the dynamical systems. Therefore, we may conclude that the Lyapunov function is decreasing along all the trajectories of the $\PWA$ dynamical systems. For more detail, please see \cite{samanipour2023stability}.
\end{proof}

The origin is assumed always to be defined as a vertex in $\Mode_0$. The assumption that the origin is always defined as a vertex of a cell ensures that we can always find a positive-definite Lyapunov function. 
Another assumption is that if a vertex $v_k\in \Vertex(X_i)$ and $v_k\in X_i\cap X_j$, then $v_k\in \Vertex(X_j)$. This assumption is required to preserve the continuity of the Lyapunov function using \eqref{eq:Equality}. Details can be found in \ref{sec:Delauney}.
%As a result of this assumption, the Lyapunov function will be continuous.    
%We should note that the proposed Lyapunov function is continuous, but not differentiable at the boundary. The stability condition of the piecewise affine function on the boundary, non-differential points, discussed in \cite{poonawala2021stability,johansson1999piecewise,samanipour2023stability} based on Filipov theorem\cite{filippov2013differential}.
\subsection{Optimization problem}\label{Relaxation}
The constraints \eqref{Eq:PD}-\eqref{eq:Equality} on variables $p_i$ and $q_i$ from \eqref{Eq:Lyap PWA} may be infeasible due to conditions (\ref{Eq:ND}) associated with the decrease of the Lyapunov function along solutions.
% search for finding $p_i$ and $q_i$ using \eqref{Eq:PD}-\eqref{eq:Equality} may not be feasible regarding the parametrization in equation \eqref{Eq:Lyap PWA}. There is a potential source of infeasibility in conditions (\ref{Eq:ND}) associated with the decrease of the Lyapunov function along solutions. 
Slack variables are added to these constraints to ensure feasibility. Consequently, we can formulate the search process for the $\PWA$ Lyapunov function as follows:
%To illustrate the ideas, consider the problem of finding a piecewise linear Lyapunov function on a polytopic partition. Rather than solving the linear programming problem as it stands in Theorem \ref{Th:General idea}, we consider the \hlctodo{red!30}{State the modification, the reader will not follow}{following slight modification}.
\begin{align}\label{eq:lyaprelaxfinal}
\min_{ p_i, q_i,\tau_i}  \quad  \sum_{i=1}^{N}\tau_i& \\
\text{Subject to:}& \nonumber\\
p_i^T (A_i v_k+a_i)-\tau_i <-\epsilon_1 \quad &\forall i \in I_1, v_k \in \Vertex(X_i)\nonumber\\
p_i^T A_i v_k-\tau_i <-\epsilon_1 \quad &\forall i \in I_0, v_k \in \Vertex(X_i)\nonumber\\
p_i^T v_k+q_i>\epsilon_2  \quad &\forall i \in I_1, v_k \in \Vertex(X_i)\nonumber\\
p_i^Tv_k>\epsilon_2  \quad &\forall i \in I_0,v_k \in \Vertex(X_i)\nonumber\\
V_i(v_k)=V_j(v_k) \quad &\forall v_k \in \Vertex(X_i)\cap \Vertex(X_j) ,\hspace{0.05em} i\neq j\nonumber\\
\tau_i\geq 0\quad & \forall i \in  \Pindex \nonumber
\end{align} 
where $\tau_{i}$ is the slack variable associated to cell $X_i$, and $\epsilon_1,\epsilon_2>0$. By design, we can state the following result. 
\begin{lemma}
The optimization problem in (\ref{eq:lyaprelaxfinal}) is always feasible.
\end{lemma}
\begin{proof}
This result is by the construction of the optimization problem.
% As a result of the construction of the optimization problem, this result is obtained. 
\end{proof}

The solution to this optimization problem yields a valid Lyapunov function if and only if all the slack variables are zero. If the cost function is non-zero, the Lyapunov function is non-decreasing at some vertices. In fact, no Lyapunov function associated with the current partition exists. It may be possible to refine the partition, meaning to divide regions within it, in order to increase the capacity of the Lyapunov function and then repeat the search using this higher-capacity function. In the following section, the refinement process is described in detail. %Multiple $\PWA$ Lyapunov functions can be used to approximate the Lyapunov function in a cell, through refinement. In particular, we divide cells that are associated with non-zero slack variables.  
\subsection{Refinement}\label{subsec:Delaunay}
% We represent cells in the partition, which are convex polytopes, by their vertices. 
% The idea here is to divide these cells into smaller ones. In order to add more flexibility to the Lyapunov function, we need to introduce a new vertex on one of the sides in the partition. Because otherwise, no flexibility is added to the Lyapunov function. For instance, if the new vertex is added to the center of the partition, then there is no refinement on the boundaries. The other important issue is that if we add a vertex to one region, we must have the same vertex on its neighbor.  \cite{johansson1999piecewise}.
% \begin{figure}
%     \centering
%     \includegraphics[scale=0.3]{Figures/Naive_approach.PNG}
%     \caption{Partitioning process in~\cite{johansson1999piecewise}. a) Two adjacent cells in 2D. b) A point on their common boundary is chosen to define the division for the cell on the left. c) The dashed red line completes the division of the cell. Continuity constraints will ensure that the Lyapunov function defines the same affine function on all three regions d) To allow the Lyapunov function to be different on the regions, the adjacent cell must also be divided using the chosen vertex.}
%     \label{fig:cell refinement Johansson}
% \end{figure}
A refinement of the current partition is intended to enhance the flexibility of the Lyapunov function search process. To achieve flexibility, a cell $X_i$ with a nonzero slack variable will be divided into smaller sub-cells. 
%After solving  the optimization problem \eqref{eq:lyaprelaxfinal}, cell $X_i$ where its slack variable $\tau_i>0$ might need a higher capacity $\PWA$. The higher capacity  
% By dividing a cell $X_i$ that has $\tau_i>0$ after solving \eqref{eq:lyaprelaxfinal} into smaller subcells, flexibility can be achieved. 
%Refinement is required if the valid Lyapunov function cannot be obtained by solving \eqref{eq:lyaprelaxfinal}. Therefore, we refine the cell $X_i$ which has $\tau_i>0$ to have smaller subcells. 
In order to keep things simple, we assume that the refinement of $X_i$ will result in the creation of two new subcells, $X_{i_1}$ and $X_{i_2}$. For each subcell, we can parameterize the $\PWA$ Lyapunov function as $V_{i_1}=p_{i_1}^Tx+q_{i_1}$ and $V_{i_2}=p_{i_2}^Tx+q_{i_2}$.
As a result, the candidate Lyapunov function for cell $X_i$ has a higher capacity $\PWA$ function that is more flexible. Furthermore, the new $\PWA$ Lyapunov function has more parameters, $p_i$ and $q_i$, as well as constraints. Increasing the number of parameters and constraints might increase the computational complexity for solving \eqref{eq:lyaprelaxfinal} since the computational complexity of linear optimization with $n$ parameters and accuracy parameter $\epsilon$ is $O(n^{3/4}log(\frac{n}{\epsilon}))$ \cite{el2012interior}. 
Therefore, to implement refinement, it is necessary to use an intelligent approach, since otherwise, the complexity of the computation may increase.

We utilize a vertex representation for the refinement process to represent cells within a partition, which are convex polytopes. The process of refinement for cells involves adding a new vertex on the cell's boundary and then forming new sub-cells. For this section, we will begin by defining a few concepts and definitions that will be useful for describing these two steps.
The first concept is the \textbf{simplex region}, a bounded region with the smallest possible number of vertices in $\mathbb{R}^d$. Polytope's faces of dimension one are called  \textbf{edges}\cite{henk2017basic}. In cell $X_i$, we define the edges by the set $\Edge(X_i)$, and each edge is represented by a pair of vertices $(v_j,v_k)$, where $v_j,v_k\in \Vertex(X_i)$. The edges of convex polytopes can be obtained by using MILP as described in \cite{zago2023vertex}. 
%each element of this set represent a pair of vertices, $(v_j,v_k)$ where $v_j,v_k\in \Vertex(X_i)$, corresponding to the edge.
It is worth emphasizing that the edges containing the origin, where $v_j=0$ or $v_k=0$, are not taken into account in the set of edges $\Edge(X_i)$. By making this assumption, we ensure that refinement will not be applied to edges containing the origin. Therefore, if $X_i\in I_0$, its subcells will always contain the origin after refinement.

%columns of $\mathbf{f}_j^k(X_i,1)$  started at $(k-1)\times d+1$ and ended $k\times d$ in the  $j^{th}$ row. As an instance, $\mathbf{f}_4^2(X_i,1)$ in $\mathcal{R}^5$ represents the columns 6 to 10 of $4^th$ row.   
% In order to distinguish between simplex and non-simple cells, we use the following set:
% \begin{align}
%     I_s=&\{i \in I(\prtition): X_i \quad \textit{is simplex}\}, \textit{and}\nonumber\\
%     I_{ns}=&\{i \in I(\prtition): X_i \quad \textit{is non-simplex}\}\nonumber,
% \end{align}
For the cell $X_i$, with dynamic $\dot x=A_ix+a_i$ and the candidate Lyapunov function $V_i=p_i^Tx+q_i$ we can define the following sets and functions.

\begin{enumerate}
\item We can find the vector field and the derivative of the Lyapunov function at a vertex, $v_j$, using the following functions.
\begin{align}
    \VectorField(X_i,v_j)=&A_iv_j+a_i,\\
    \derivative(X_i,v_j)=&p_i^T\VectorField(X_i,v_j),
\end{align}
where $\VectorField(X_i,v_j)\in \mathbb{R}^n$ is the vector field of the local dynamic at the vertex $v_j$, and $\derivative(X_i,v_j)$ is the derivative of the Lyapunov function at the specified vertex in $X_i$.
%     \item The following index set can be used to determine the indices of vertices on each edge of a cell:
% \begin{align}\label{eq:side index}
%     I_{edge}(i)=&\{(j,k) \in \Vertex(i): \alpha v_j+(1-\alpha) v_k \in \partial X_i \nonumber\\
%     & \forall \quad 0\leq\alpha\leq 1,j\neq k, v_j\neq 0,v_k\neq 0  \}
% \end{align}
%It is important to note that edges containing the origin are not included in the index set.
%\item For the indices of the vertices on the longest edge of a cell, the following function can be used:
% \begin{align}\label{eq:longest_side}
%     I_{Length}(i)= &\{(j,k)\colon \arg \max_{j,k} (\lvert v_j-v_k \rvert_2) \nonumber\\
%     &\quad \forall j,k \in I_{edge}(i)\}.
% \end{align}
\item The vertices of the longest edge of a cell can be obtained using the following function:
\begin{align}\label{eq:longest_side}
    L_{max}(i)= &\argmax_{(v_j,v_k)\in \Edge(X_i)} (\lvert v_j-v_k \rvert_2)
\end{align} 
%\end{enumerate}
%Now Let's consider the cell $X_i$, with dynamic $\dot x=A_ix+a_i$ and the candidate Lyapunov function $V_i=p_i^Tx+q_i$. Then, we can define the following sets and functions.
%\begin{enumerate}
\item The following function can be used to capture changes in the sign of the derivative of a candidate Lyapunov function:
% \begin{align}\label{eq:sign Vdot}
%     \operatorname{\dot V}_{sgn}(i)= \begin{cases}1 &\text { if } \operatorname{sgn}\left(p_i^T\left(A_i v_k+a_i\right)\right) \geq 0\\ & \forall k \in \mathcal{V}(i) \\ -1 &\text { otherwise, }\end{cases}
% \end{align}
\begin{align}\label{eq:sign Vdot}
% sgn_{\partial V}
    \SignFunction(i)= \begin{cases}1 & \operatorname{sgn}(\derivative(X_i,v_j)) \geq 0, \forall v_j \in \Vertex(X_i),\\ 
    -1 &\operatorname{sgn}(\derivative(X_i,v_j)) \leq 0, \forall v_j \in \Vertex(X_i),\\ 
    0 &\text { otherwise, }\end{cases}
\end{align}
 where the $\operatorname{sgn(x)}$ is the standard sign function. This function generates zero whenever the sign of the derivative of the candidate Lyapunov function in the cell $X_i$ changes. Otherwise, this function generates either 1 or -1 depending on the sign of the derivative of the candidate Lyapunov function within the cell $X_i$.
\item With $\SignFunction(i)$ being 0, the following set may also be used to provide the vertices of the edges where the sign of the derivative of the Lyapunov function has changed.
\begin{align}\label{eq:inflection point}
    &c_{V}(i)=\{(v_j,v_k) \colon \SignFunction(i)=0, \forall (v_j,v_k) \in \Edge(X_i),\nonumber \\ 
    &\derivative(X_i,v_j)\derivative(X_i,v_k)<0\}.
    %& (p_i^T(A_iv_j+a_i)) (p_i^T(A_iv_k+a_i))<0\}.
    %indx= \underset{k,j}{\mathrm{argmax(|p_i^TA_i(v_k-v_j))|}}\quad \forall %k,j \in I_{side}
\end{align}
There are multiple edges where the sign of the derivative of the Lyapunov function has changed if $\SignFunction(i)=0$. 
\item The following equation can be used to determine the vertices for an edge with the largest variation in the derivative of a candidate Lyapunov function.
\begin{align}\label{eq:Biggest Vdot variation}
    & \Delta\dot V_{max}(i)= \{(v_j,v_k)\colon %(v_j,v_k) \in \Edge(X_i),  
    \\
    &\argmax_{(v_j,v_k)\in \Edge(X_i)}|\derivative(X_i,v_j)-\derivative(X_i,v_k)|\}.\nonumber
\end{align}
\item We aim to determine the edge along which the vector fields exhibit the greatest range of angle variations. Therefore, we use the following function to find the edge with the smallest cosine between the vector field at its vertices. 
\begin{align}\label{eq:smallest cosine}
     &cos_{min}(i)= \{(v_j,v_k)\colon\\
    &\underset{(v_j,v_k)\in \Edge(X_i)}{\arg \min} \frac{\langle \VectorField(X_i,v_j),\VectorField(X_i,v_k)\rangle}{|\VectorField(X_i,v_j)|_2|\VectorField(X_i,v_k)|_2}\}\nonumber 
\end{align}
\end{enumerate}

 Now, we can delve into the refinement process. 
\subsubsection{Finding new vertices}\label{sec:new vertices}
The first step in the refinement process is to introduce new vertices on the boundary of cells in $I_s=\{i \in I \colon \tau_i>0\}$. The new vertex for the cell $X_i$ can be obtained using the following equation:
        \begin{equation}\label{eq:new vertex}
         v_{new_i}=\alpha v_j+\beta v_k,    
        \end{equation}
which is a linear combination of two vertices of an edge. Based on the splitting approach which will be introduced in this section, $\alpha,\beta$, $v_j$, and $v_k$  in \eqref{eq:new vertex} could be different.
Here we consider three different approaches for finding the new vertices. 
\paragraph{Naive refinement}
The first algorithm is inspired from \cite{johansson1999piecewise}. The original algorithm was described for simplex regions and restricted to 2-D problems. In order to make the comparison possible we generalize the method for all types of regions. The process of refinement for the cell $X_i$ based on \cite{johansson1999piecewise} is described in \textbf{Algorithm \ref{alg:naive algorithm}}.
\begin{algorithm}[tb]
    \begin{algorithmic}
    \REQUIRE  $\PWA(x)$, Vertices  
    \STATE{1- Find
    \begin{equation}\label{eq:Biggest slack variable}
        i=\argmax_{i\in I_s} (\tau_i)\
    \end{equation}}
    \STATE{2- Find $(v_j,v_k)$ using $L_{max}(i)$ \eqref{eq:longest_side}}.
    \STATE{3- finding $v_{new_i}$ using \eqref{eq:new vertex} where $\alpha=\beta=0.5$.}
    \RETURN $v_{new_{i}},i$.
    \end{algorithmic}
    \vspace{1em}
    \caption{Finding new vertices using naive refinement}
    \label{alg:naive algorithm}
\end{algorithm} 
The naive algorithm adds a new vertex exclusively to the longest edge, denoted as $L_{max}$, of cell $X_i$ that has the largest slack variable, as determined by \eqref{eq:Biggest slack variable}. This method creates sub-cells with the largest possible volume without considering the candidate Lyapunov function or local dynamics. Consequently, it may lead to unsatisfactory results. 
Selecting vertices randomly could increase computational complexity without necessarily improving the refinement process. Thus, it is crucial to choose new vertices intelligently. 
\paragraph{Lyapunov-based refinement} 
To address the challenge with the naive refinement, a new approach is proposed, leveraging the candidate Lyapunov function to make more informed decisions regarding selecting new vertices. The basic principle behind this method is that for every cell $X_i$ where $i\in I_s$ finding a set of points $P(i)=\{v_{new_i}\colon \derivative(X_i,v_{new_i})=0, v_{new_i}\in \partial X_i\}$. 
%$v_{new_i}$ on the edges obtained from $c_v(i)$ where $\dot V_i(v_{new_i})=0$. 
Using these points, $v_{new_i}\in P(i)$, the cell $X_i$ could be divided into the two sub-cells, $X_{i_1}$ and $X_{i_2}$, where $\SignFunction(i_2)=-\SignFunction(i_1)$. In the case of $\SignFunction(i)=0$, we know that $P(i)\neq \emptyset$.   
%Given the optimization problem (\ref{eq:lyaprelaxfinal}), $X_i$ with nonzero slack variables has at least one vertex, $v_1$, where $\dot V_i(v_1)>0$. 
%In order to solve this problem the Lyapunov-based refinement is proposed. 
%The principle behind this technique is to separate vertices on a side with different signs of the $\dot V$. If there is no change in the sign of $\dot V$, then we separate the vertices on a side with the biggest difference in the $\dot V$ to split. 
%As a result, we select the new vertex on the side where $\dot V$ has the greatest difference. using convex optimization. Furthermore, if the sign of the $\dot V$ has not changed, then we choose the midpoint of the side with the largest difference in $\dot V$.
%It is essential to note that we need to add the new vertex on the side, as adding it to the interior of a cell will not increase the flexibility of the boundary \cite{johansson1999piecewise}. 
Therefore, we can find these points using the following convex problem in cell $X_i$.
\begin{align}\label{eq:finding zero Lyap}
        & \max_{\alpha,\beta} \quad 0\\
        &\textit{s.t.}\quad \alpha \derivative(X_i,v_j)+\beta \derivative(X_i,v_k)=0,\nonumber\\
        & \alpha+\beta=1,\nonumber\\
        & 0\leq\alpha,\beta\leq1,\nonumber\\
        & (v_j,v_k) \in \Edge(X_i). \nonumber
\end{align}
An explanation of how to find $i$, $v_j$ and $v_k$ in \eqref{eq:finding zero Lyap} and other details about finding a new vertex using Lyapunov-based refinement can be found in \textbf{Algorithm} \ref{alg:Lyapunov-based algorithm}.
\begin{algorithm}[tb]
    \begin{algorithmic}
    \REQUIRE  $\PWA(x)$, Vertices  
    \FOR{$i \in I_s$} \STATE{1- Finding $sgn_{\dot V}(i)$ using \eqref{eq:sign Vdot}. 
    \IF{$sgn_v(i)=0$}
    \FOR{$(v_j,v_k) \in c_V(i)$} {\STATE {1- Solve the convex problem \eqref{eq:finding zero Lyap} to obtain $\alpha$ and $\beta$}.\STATE {2- Find $v_{new_i}$ using \eqref{eq:new vertex}.}
        }\ENDFOR
    \ELSE
        \STATE {1- Find $(v_j,v_k)$ using $\Delta\dot V_{max}(i)$ \eqref{eq:Biggest Vdot variation}}.
        \STATE {2- Find $v_{new_i}$ using \eqref{eq:new vertex} where $\alpha=\beta=0.5$.}        
    \ENDIF
    }\ENDFOR
    \RETURN $v_{new_{i}},i$.
    \end{algorithmic}
    \vspace{1em}
    \caption{Finding new vertices using Lyapunov-based refinement}
    \label{alg:Lyapunov-based algorithm}
\end{algorithm}
% \begin{align*}
%     X_{i1}=\{x \in X_i\colon \dot V_i(x)\geq 0\},\\
%     X_{i2}=\{x \in X_i\colon \dot V_i(x)\leq 0\}.
% \end{align*}
%by finding new vertices $v_{new_{i}}$ where $\dot V_i(v_{new_{i}})=0$ using \eqref{eq:finding zero Lyap}.
If $\SignFunction(i)=1$, then $P(i)=\emptyset$, so we choose the new vertex at the edge obtained from $\Delta\dot V_{max}(i)$.

\begin{figure}
    \centering
    \includegraphics[scale=0.32]{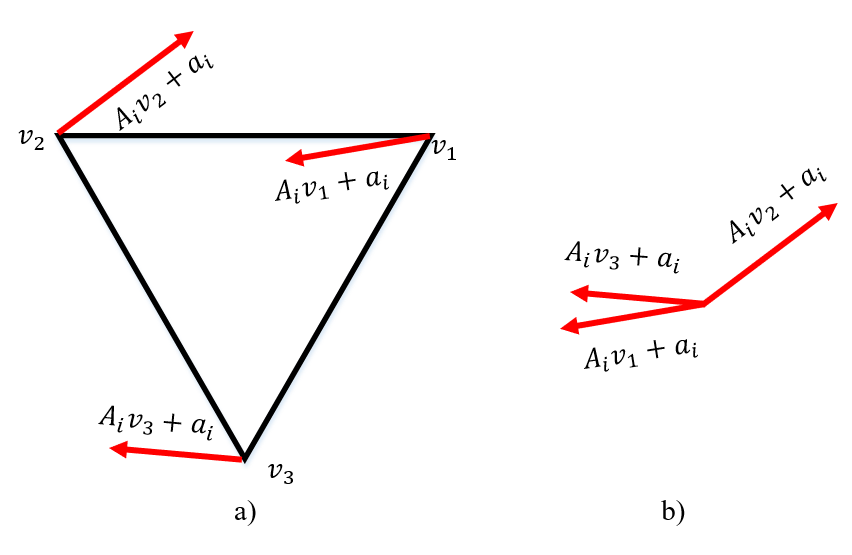}
    \caption{Vector fields of a cell on its vertices. b) The angle between vector fields of vertices. As can be seen, the angle of the vector field between $v_1$ and $v_2$ is close to $-\pi$.}
    \label{fig:vector_field}
\end{figure}

In contrast to the previous method that focused only on the cell with the largest slack variable, the Lyapunov-based refinement is now applied to all cells with nonzero slack variables, denoted as $i \in I_s$. As a result of this broader approach, each relevant cell will be refined based on its individual candidate Lyapunov function. 
%Additionally, new vertices are selected based on the candidate Lyapunov function may facilitate the search process for the next iteration. 
However, it is important to note that the coefficient vector $p_i$ used in the refinement process may change significantly in the next iteration. Therefore, this approach may not be suitable in all cases, as the optimization process in the subsequent steps can alter the candidate Lyapunov function.
\paragraph{Vector field refinement}
To address the problem with the Lyapunov-based refinement, the search for new vertices should be conducted using a method that is not influenced by the optimization process in the subsequent steps.
The proposed method leverages the vector field information of the dynamics, which remains unchanged during the optimization process.
The underlying heuristic behind this method is that the direction or magnitude of the vector fields along an edge may undergo significant changes within a cell $X_i$ where $i \in I_s$. Consequently, a higher-capacity PWA function may be required to represent the Lyapunov function within $X_i$ accurately.
As illustrated in Fig.\ref{fig:vector_field}, the vector field direction in a cell can exhibit substantial variations, such as a flip from $v_1$ to $v_2$. In such cases, a simple $\PWA$ function may struggle to approximate the level set accurately. To mitigate this, the method is adding a new vertex, $v_{new_i}$, between $(v_1,v_2)$ where $\angle(\VectorField(X_i,v_1),\VectorField(X_i,v_{new_i}))=\angle(\VectorField(X_i,v_2),\VectorField(X_i,v_{new_i}))$.
% the method aims to smooth the variation in the vector field along the edge $(v_1, v_2)$ by introducing a new vertex between its existing vertices. This new vertex is located at the angle bisector of the vector fields.
Consequently, after each refinement process, the greatest angle between the vector fields of an edge in cell $X_i$ is divided in half.
The process of finding a new vertex using the vector field refinement is outlined in \textbf{Algorithm} \ref{alg:Vector-field  algorithm}, which provides a detailed description of the method.
\begin{algorithm}[tb]
    \begin{algorithmic}
    \REQUIRE  $\PWA(x)$, Vertices  
    \FOR{$i \in I_s$} \STATE{1- Find $(v_j,v_k)$ using \eqref{eq:smallest cosine}. \STATE {2- Find $\alpha$ and $\beta$ using following equation:
    \begin{equation}\label{eq:Angel bisector}
        \alpha=\frac{1}{1+\frac{|\VectorField(X_i,v_j)|_2}{|\VectorField(X_i,v_k)|_2}},
        \beta=1-\alpha.
    \end{equation}}
    \STATE{3- Find the new vertex using \eqref{eq:new vertex}}
    }\ENDFOR
    \RETURN $v_{new_{i}},i$.
    \end{algorithmic}
    \vspace{1em}
    \caption{Finding new vertices using Vector field refinement}
    \label{alg:Vector-field  algorithm}
\end{algorithm}
%As a result, a new refinement method, vector field refinement is introduced. 
%the first step in this method is to determine the angle between the vector fields of each connected vertex. In the following step, we will find the largest angle and then separate the two vertices associated with the biggest angle. Finally, we will choose the new vertex between $v_j$ and $v_k$ corresponding to the biggest angle using the following equation:
%we will be able to locate the new vertex where its vector field is the angle bisector. Hence, during each optimization iteration, the greatest angle between the vector fields will be divided in half. 

Before moving on to the next step, storing the new vertices created by these algorithms in the following buffer is necessary.
\begin{equation}\label{eq:Buffer}
    B=\{v_{new_i} \in \mathbb{R}^n \colon i \in I_s\}.
\end{equation}
Now we can proceed to the next step, which is forming sub-cells.  
\subsubsection{Forming sub-cells}\label{sec:Delauney}
In order to form sub-cells, Johannson\cite{johansson1999piecewise} proposed remedies for 2-D systems; however, this method is limited to simplex cells. It was suggested that triangulation methods be used for non-simplex regions in \cite{johansson1999piecewise}, but no specific method or implementation is presented. It has also been proposed in \cite{rubagotti2011stability} to apply Delaunay triangulation to all cells; however, the results have been limited to 2-D examples. 
We apply Delaunay triangulation to overcome the challenges associated with forming sub-cells for non-simplex cells and cells in higher dimensions ($n>2$), which would be challenging to accomplish manually.

\begin{figure}
    \centering
    \includegraphics[scale=0.5]{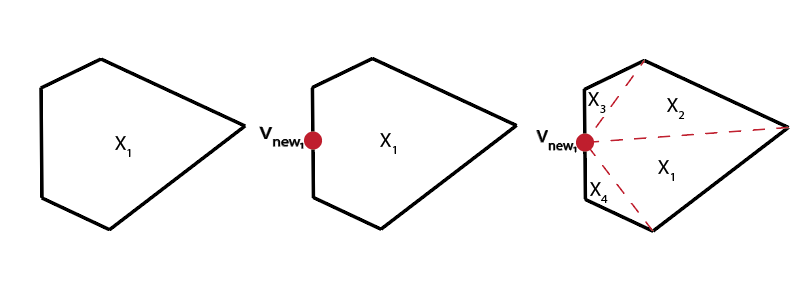}
    \caption{Triangulation process using Delaunay. a) A non-simplex region b) Midpoint is chosen to be added as a new vertex. c) The refined cells generated by the Delaunay triangulation.}
    \label{fig:Delaunay process}
\end{figure}

The Delaunay triangulation of a set of points in $\mathbb{R}^d$ is defined to be the triangulation such that the circumcircle of every triangle in the triangulation contains no point from the set in its interior. Such a unique triangulation exists for every point set in $\mathbb{R}^d$, and it is the dual of the Voronoi diagram. 
Moreover, the Delaunay triangulation will maximize the minimum angle in each triangle\cite{rajan1991optimality}. $DT(\Vertex(X_i))$ is the notation for implementing Delaunay triangulation using the vertices of the cell $X_i$. The process of implementing Delaunay triangulation for a single cell is illustrated in Fig.\ref{fig:Delaunay process}.
Delaunay triangulation will also handle the continuity of the Lyapunov function if the partition is composed of multiple cells. 
To illustrate how continuity is preserved, let us consider the $v_{new_i}$ as the new vertex obtained using \eqref{eq:new vertex} for the cell $X_i$. If $v_{new_i}\in X_i\cap X_j$, then $v_{new_i}$ must also be considered as a new vertex for the cell $X_j$, and $DT(\Vertex(X_i)\cup v_{new_i})$ and $DT(\Vertex(X_j)\cup v_{new_i})$ should be implemented. Consequently, even after refinement, continuity would be guaranteed by \eqref{eq:Equality}. Generally, in order to implement Delaunay triangulation within the current partition, we have to follow the following steps.
\begin{enumerate}
    \item First, we must obtain the following set containing cells that required refinement.
\begin{align}
    I_{split}=&\{i: X_i \cap  B\neq \emptyset, i \in I(\prtition)\}.
\end{align}
\item Then, we need to find the vertices located on the boundary of the cell $X_i$ where $i\in I_{split}$ using the following set.
\begin{align}
    \mathcal{V}_{new}(i)=&\{v_{new_j}\colon v_{new_j}=X_i\cap B, i\in I_{split}\}.
\end{align}
\item Then, we can form the new sub-cells using $DT(\Vertex(X_i)\cup \mathcal{V}_{new}(i))$ for $i \in I_{split}$.
\end{enumerate}

\begin{figure}
    \centering
    \includegraphics[scale=0.3]{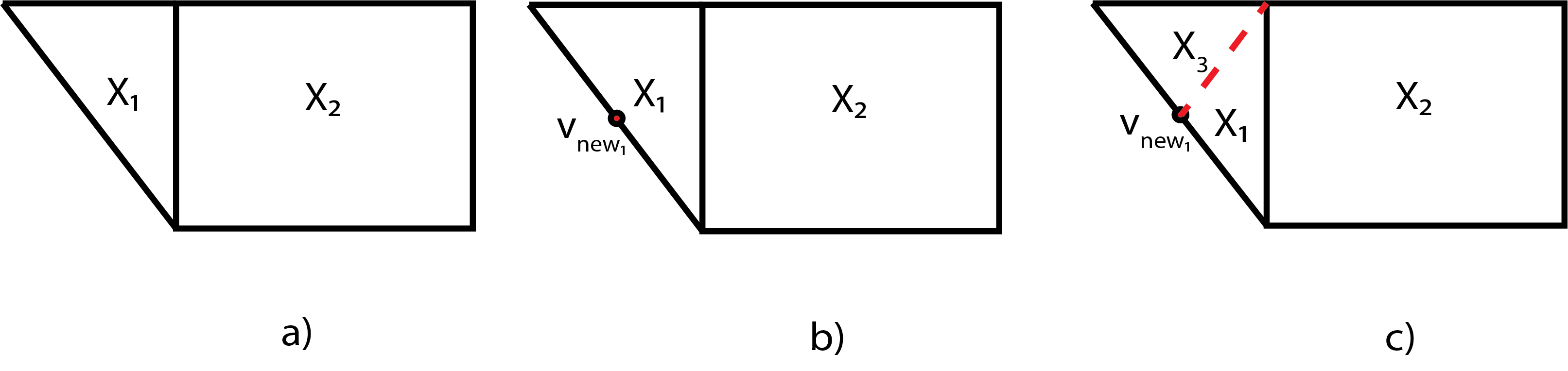}
    \caption{The process of refinement using naive refinement  a) Two adjacent cells where $\tau_1>\tau_2>0$. b) Based on the naive refinement method, the new vertex is located on the longest edge of the cell $X_1$ because it has the greatest slack variable $\tau_1$. c) Delaunay triangulation is used to form the sub-cells.}
    \label{fig:naive approach}
\end{figure}

\begin{figure}
    \centering
    \includegraphics[scale=0.3]{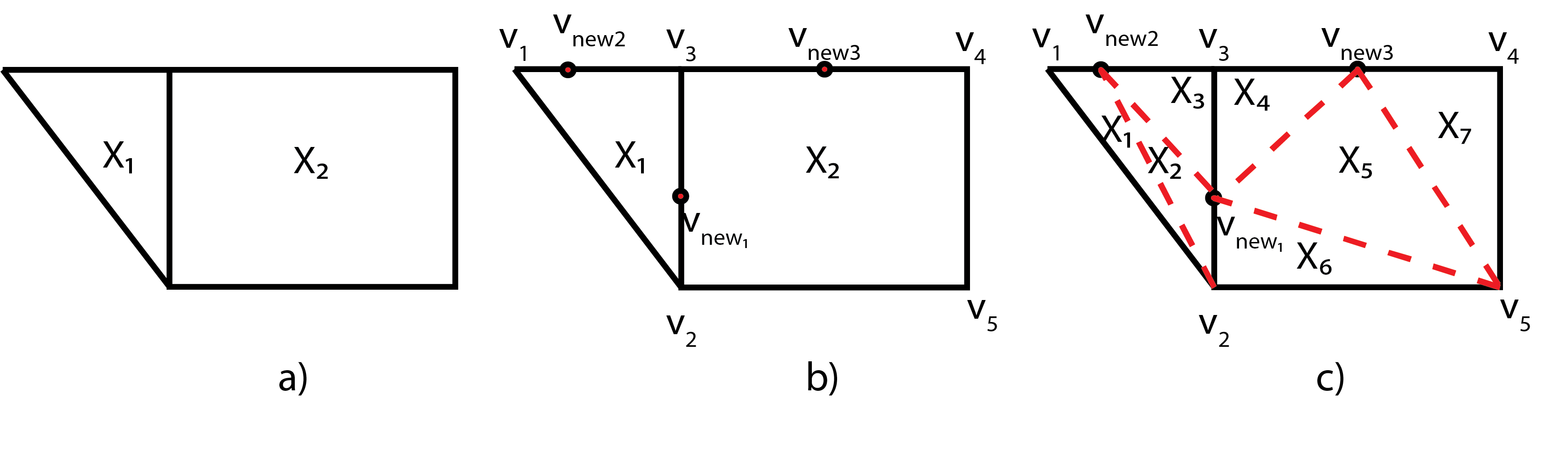}
    \caption{The process of refinement using Lyapunov-based refinement  a) Two adjacent cells where $\tau_1>\tau_2>0$. b) Let's assume in the simplex cell, $X_1$, we have $\derivative(X_1,v_2)<0<\derivative(X_1,v_1)<\derivative(X_1,v_3)$ and in the non-simplex-cell, $X_2$, $0<\derivative(X_2,v_3)<\derivative(X_2,v_2)<\derivative(X_2,v_5)<\derivative(X_2,v_4)$. Based on the Lyapunov-based refinement method, the new vertices, $V_{new_1}$ and $V_{new_2}$, for the simplex cell will be obtained using \eqref{eq:finding zero Lyap} on the edges obtained using \eqref{eq:inflection point}. For the non-simplex cell, the new vertex $V_{new_3}$ is obtained using\eqref{eq:new vertex} where $\alpha=\beta=0.5$ on the edge obtained using \eqref{eq:Biggest Vdot variation} c) Delaunay triangulation is used to form the sub-cells for the cell $X_1$ and $X_2$ with their new vertices. Delaunay Triangulation for the cell $X_1$ and $X_2$ will be $DT(\Vertex(X_1)\cup v_{new_1}\cup v_{new_2})$ and $DT(\Vertex(X_2)\cup v_{new_1}\cup v_{new_3})$ respectively.}
    \label{fig:Lyapunov based}
\end{figure}

\begin{figure}
    \centering
    \includegraphics[scale=0.30]{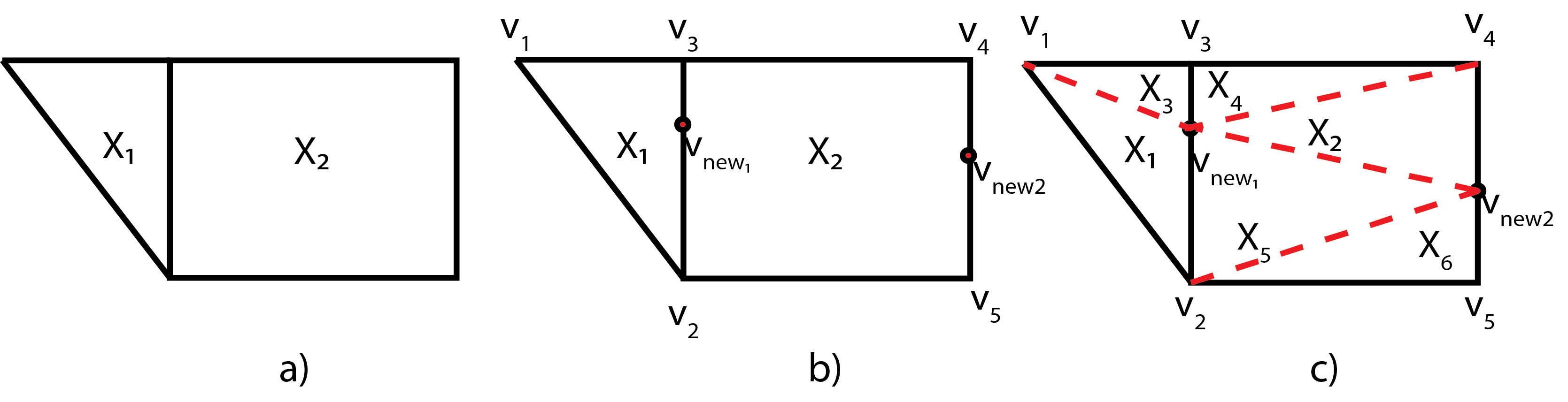}
    \caption{The process of refinement using vector field refinement  a) Two adjacent cells where $\tau_1>\tau_2>0$. b) Let's assume in the simplex cell we have the biggest variation in the vector field angle, $cos_{min}(1)=(v_2,v_3)$, and in the non-simplex cell the biggest variation of the vector field, $cos_{min}(2)=(v_4,v_5)$. Based on the vector field refinement method, the new vertex for the simplex and non-simplex cell will be obtained using \eqref{eq:new vertex} with $\alpha$ and $\beta$ obtained from \eqref{eq:Angel bisector}. c) The Delaunay triangulation is employed to form the sub-cells for the cell $X_1$  and $X_2$ using $DT(\Vertex(1)\cup v_{new_1})$ and $DT(\Vertex(2)\cup v_{new_1}\cup v_{new_2})$ respectively.}
    %with its new vertex $V_{new_1}$. Considering the fact that $V_{new_1}$ is on the common edge of two cells, it is also considered to be a new vertex for the cell $X_2$. Delaunay Triangulation creates subcells for cell $X_2$ with these two new vertices.}
    \label{fig:Vector_field}
\end{figure}

%To describe the process we define the following 
%ensure the continuity of the Lyapunov function,
% , we define an index set that includes all the cells with one or more new vertices on their boundary as follows:
% \begin{align}
%     &I_{split}=\{i: X_i \cap  {\textit{buffer}}_{v}\neq \emptyset, i \in I(\prtition)\},\\
%     &I_{v_{new}}(i)=\{v_{new_j}\in X_i \cap  {\textit{buffer}}_{v}: i\in I_{split}\}.
% \end{align}
% We can determine whether a cell $X_i$ needs to be split by using $I_{split}$ and if so, we can find the new vertices on the boundary of $X_i$ using $I_{v_{new}}(i)$. Finally, we can use the Delaunay triangulation for cell $X_i$ where $i\in I_{split}$ to split it with all the new vertices $I_{v_{new}}(i)$. By using this approach, if the new vertex $v_{new_i}$ obtained using the proposed methods for the cell $X_i$ is also in $X_i\cap X_j$, it will be counted as the new vertex of $X_j$. Consequently, even after refinement, continuity would be guaranteed by \eqref{eq:Equality}\todo{I am not sure if it is good enough}. 
The process of refinement based on the naive approach using Delaunay triangulation is shown in Fig. \ref{fig:naive approach}. As can be seen, the sub-cells are created just in the simplex cells. However, the Lyapunov-based and vector-field methods perform differently as shown in Fig.\ref{fig:Lyapunov based}. and Fig.\ref{fig:Vector_field}. respectively. 
\section{Results}\label{sec:res}
The paper presents seven examples to demonstrate the search performance for a $\PWA$ Lyapunov function using the algorithm described in \textbf{Algorithm} \ref{alg:refinesearch}. The computations are implemented using the Mosek optimization package \cite{aps2022mosek} and Python 3.9 on a computer with a 2.1 GHz processor and 8 GB RAM.
During the computations, a tolerance of $10^{-8}$ is used to determine if a number is nonzero. In all the examples, the values of $\epsilon_1$ and $\epsilon_2$ are set to $10^{-4}$.
These examples aim to showcase the effectiveness and efficiency of the proposed algorithm in finding valid Lyapunov functions within reasonable computation times.
\begin{example}[4-D Example \cite{chen2021learning}]
\label{example:4D}
For this example, we will use the 4-D MPC example presented in \cite{chen2021learning} as follows:
\begin{align}\label{eq:MPC4D}
    x_{t+1}=&\begin{bmatrix}
      0.4346&-0.2313&-0.6404&0.3405\\
-0.6731&0.1045&-0.0613&0.3400\\
-0.0568&0.7065&-0.086&0.0159\\
0.3511 &0.1404&0.2980&1.0416
  \end{bmatrix}x_t+\\
  &\begin{bmatrix}
      0.4346,-0.6731,-0.0568,0.3511
  \end{bmatrix}u_t.\nonumber
\end{align}
It includes the same details as \cite{chen2021learning}, such as a state constraint of $\lVert x \rVert_{\infty} \leq 4$, an input constraint of $\lVert u \rVert_{\infty} \leq 1$, a prediction horizon of $T=10$, a stage cost of $Q=10I$ and $R=1$. 
%With the state constraint $\lVert x \rVert_{\infty} \leq 4$ and the input constraint $\lVert u \rVert_{\infty} \leq 1$ we design the MPC controller by choosing horizon $T=10$, stage cost with $Q=10I$ and $R=1$, terminal cost with $P_\infty$, the solution to Ricatti equation defined by $(A,B,Q,R)$, and the terminal set is chosen to maximize positive invariant set.
Explicit MPC produces a $\PWA$ dynamic with 193 cells. To ensure that the origin is a vertex, we refined the cell with the origin on its interior first. Our next step is to convert the discrete-time $\PWA$ dynamics into continuous-time $\PWA$ dynamics with a sampling time $t_s=0.01$. Finally, We searched for the continuous $\PWA$ Lyapunov function using \textbf{Algorithm} \ref{alg:refinesearch} with all refinement techniques.  The \textbf{Algorithm} \ref{alg:refinesearch} timed out after 2000 seconds using the naive refinement after 31 iterations. 
\textbf{Algorithm} \ref{alg:refinesearch} found the Lyapunov function in 1200 seconds using the Lyapunov-based refinement with 5874. With the vector field refinement, the \textbf{Algorithm} \ref{alg:refinesearch} found the solution in 280 seconds by generating 3086 cells.
In comparison with \cite{chen2021learning}, the Lyapunov function using vector-field refinement requires a shorter computational time.
\end{example}
\begin{example}[4-D controllable canonical dynamic]
\label{example:4D canonical}
Following is a simple 4-D example with stable canonical controllable dynamics with condition number 10 to illustrate the meaningful difference between the refinement methods.
\begin{align}\label{eq:4D-Canonical}
    \dot x=&\begin{bmatrix}
      0&1&0&0\\
0&0&1&0\\
0&0&0&0\\
-24 &-50&-35&-10
  \end{bmatrix}x,
\end{align}
 where $\lVert x \rVert_{\infty} \leq 5$ and the initial partition includes 16 simplex cells around the origin with the dynamic \eqref{eq:4D-Canonical}. The search \textbf{Algorithm} \ref{alg:refinesearch} found the valid $\PWA$ Lyapunov function after 43 seconds with 1054 cells created as a result of vector field refinement, whereas Lyapunov-based refinement required 106 seconds with 2743 cells, and naive search required 1546 seconds with 6943 cells.   
 \end{example}
\begin{example}[Path Following Wheeled Vehicle\cite{zhou2022neural}]
\label{example:pathfollowing}
The following kinematic model is used to analyze the stability of a path following wheeled vehicle in \cite{zhou2022neural}:
\begin{align}\label{eq:path following}
&\dot{d}_e=\nu sin(\theta_e),\\
&\dot{\theta}_e=\omega-\frac{\nu \kappa(s) cos(\theta_e)}{1-d_e\kappa(s)}.\nonumber
\end{align}
\begin{figure}
    \centering
    \includegraphics[scale=0.48]{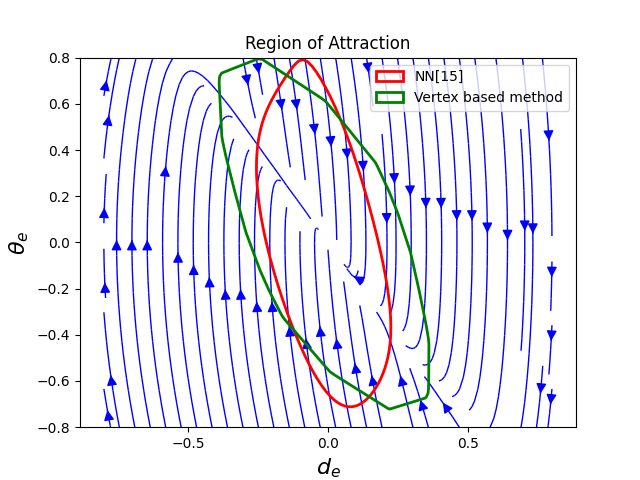}
    \caption{The ROA for the closed-loop path following wheeled vehicle using NN\cite{zhou2022neural} and vertex-based method.}
    \label{fig:pathfollowing}
\end{figure}
In equation (\ref{eq:path following}), we have the state variables $\theta_e$, which represents the angle error, and $d_e$, which represents the distance error. The control input is denoted as $\omega$. 
%We employed the neural network (NN) controller proposed in \cite{zhou2022neural}. 
%We assumed that the desired path is a unit circle, where the curvature $\kappa(s)$ is equal to 1.
In this study, we used a single-hidden layer ReLU with 50 neurons as described in \cite{samanipour2023stability} in order to identify the dynamic \eqref{eq:path following} with the NN controller \cite{zhou2022neural} in the region $\lVert x \rVert_{\infty} \leq 0.8$. Moreover, we used the vertex-based method along with vector field refinement to obtain the $\PWA$ Lyapunov function. As can be seen in Fig. \ref{fig:pathfollowing},
a comparison was made between the ROA obtained by the proposed method and the ROA obtained using the NN Lyapunov function \cite{zhou2022neural}.
%Our results indicate that the NN Lyapunov function derived through our method is less conservative compared to the NN Lyapunov function derived in Zhou et al. (2022) \cite{zhou2022neural}.
% In (\ref{eq:pf}), state variables are $\theta_e$, the angle error, and $d_e$, the distance error, and the control is $\omega$. We used the NN controller designed in \cite{zhou2022neural}, and assumed that the target path is a unit circle, $\kappa(s)=1$. We identified the closed-loop dynamic within $\lVert x \rVert_{\infty} \leq0.8$ with 50 neurons using single-hidden layer ReLU. As a result of applying the vector field refinement method to the identified dynamics, we obtained the Lyapunov function. Fig.~\ref{fig:pathfollowing} illustrates the vector fields and the region of attraction. Based on our results, the NN Lyapunov function derived by our method is less conservative than the NN Lyapunov function derived by ~\cite{zhou2022neural}. 
\end{example}
\begin{example}[Multi-agent consensus]
The Hegselmann-Krause model is a widely studied model in the literature, which involves $N$ autonomous agents with state variables $\xi_i$. Each agent's dynamics are given by the equation:
\begin{equation}
\dot{\xi_i} = \sum_{j=1}^{N} \phi(\xi_i,\xi_j)(\xi_j-\xi_i)
\end{equation}
where $i$ ranges from $1$ to $N$, and $\phi:[0,1]^2\rightarrow \{0,1\}$ represents a weight function as defined in the reference \cite{iervolino2017lyapunov}.
The stability analysis results for this model are presented in Fig. \ref{fig:Multi_agent}. We observed that a valid $\PWA$ Lyapunov function can be obtained without requiring any refinement. Therefore, the choice of different splitting approaches does not have any impact on this particular example. The details are provided in Table \ref{table:summary}.
\label{example:Multi-agent}
\begin{figure}
    \centering
    \includegraphics[scale=0.48]{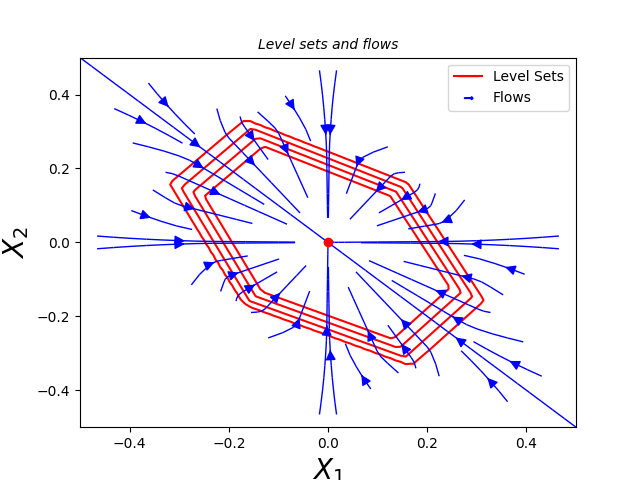}
    \caption{Selected level sets and flows for the Example \ref{example:Multi-agent}.}
    \label{fig:Multi_agent}
\end{figure}
\end{example}
\begin{example}[2-D example from \cite{della2019smooth},\cite{poonawala2021stability},\cite{johansson1999piecewise}]
This system has been presented in four different regions as follows:
\begin{align}
    &Z_1=\{x\in \mathbb{R}^2: -x_1+x_2\geq 0, x_1+x_2\geq 0\} \\
    & Z_2=\{x\in \mathbb{R}^2: -x_1+x_2\geq 0, -x_1-x_2\geq 0\}\nonumber\\
    &Z_3=\{x\in \mathbb{R}^2: x_1-x_2\geq 0, -x_1-x_2\geq 0\}\nonumber\\
    &Z_4=\{x\in \mathbb{R}^2: x_1-x_2\geq 0, x_1+x_2\geq 0\}\nonumber
\end{align}
and the dynamics are as follows:
\begin{align}
    \Omega_p:\dot{x}=\left\{
   \begin{array}{lr}
        \begin{bmatrix}
             -0.1& 1 \\
            -5 & -0.1 
        \end{bmatrix}x &\quad if \hspace{0.3em} x\in Z_1 \hspace{0.3em} or\hspace{0.3em} x\in Z_3 \\
        \begin{bmatrix}
             -0.1& 5 \\
            -1 & -0.1 
        \end{bmatrix}x &\hspace{0.3em} if \hspace{0.3em} x\in Z_2 \hspace{0.3em} or\hspace{0.3em} x\in Z_4.
   \end{array}
   \right.
\end{align}
The level sets and the vector fields are shown in Fig.\ref{fig:Flower_like dynamic}. The $\PWA$ Lyapunov function was obtained by refining the cells. In this example, all three refinement methods perform similarly in finding the Lyapunov function. The details about this example are presented in Table\ref{table:summary}. The refinement process creates 128 cells in the partition.
%These level sets can be obtained by PWQ Lyapunov function without refinement \cite{johansson1999piecewise}.  
\label{example:Flower_like dynamic}
\begin{figure}
    \centering
    \includegraphics[scale=0.48]{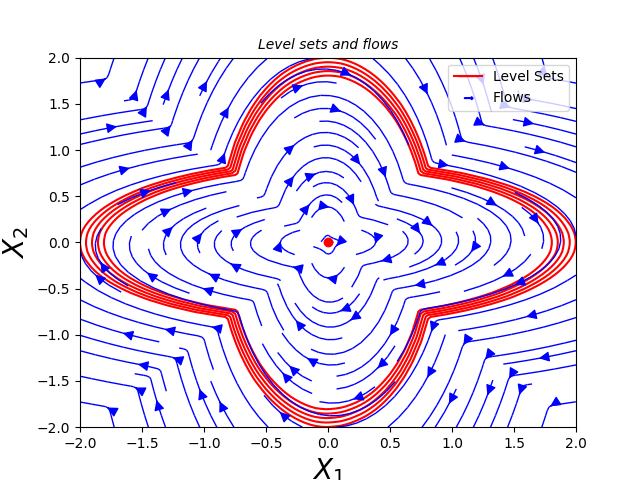}
    \caption{Selected level sets and flows of the Example \ref{example:Flower_like dynamic}.}
    \label{fig:Flower_like dynamic}
\end{figure}
\end{example}
\begin{example}[Explicit model-predictive controller\cite{samanipour2023stability}]
In this study, the stability of the following discrete time dynamic is investigated using explicit MPC, similar to \cite{chen2021learning,samanipour2023stability}. 
%Model Predictive Control is used to derivate a control policy for a discrete-time linear dynamical system. Similar to \cite{samanipour2023stability}, we consider the following dynamics. 
\begin{align}
  x_{t+1} = \begin{bmatrix}
        1 & 1\\
        0 & 1 
\end{bmatrix}x_{t}+\begin{bmatrix}
        1 \\
        0.5
\end{bmatrix}u_{t}\label{eq:mpcdiscrete}
\end{align}
As in \cite{samanipour2023stability}, the MPC problem has the same specification such as stage cost, actuator, and state constraints. 
We use the {\tt MPT3} toolbox \cite{kvasnica2004multi} in {\tt Matlab} to obtain an explicit controller. A sampling time of $t_s=0.01s$ was used to obtain the continuous form of the dynamic \eqref{eq:mpcdiscrete} with the explicit MPC controller.
%the closed-loop continuous-time dynamics with the Explicit MPC controller.   
%The performance of this explicit MPC controller has been verified on the continuous time dynamical system. 
The $\PWA$ dynamics generated by the explicit MPC have a cell where the origin is not on the vertices. As a result, we refine this cell with the origin as a new vertex, $DT(\Vertex(X_i)\cup 0)$, and then start the \textbf{Algorithm} \ref{alg:refinesearch}. 
% \begin{align}
%   \dot x = \begin{bmatrix}
%   0 &100\\0 & 0
%   \end{bmatrix}x + \begin{bmatrix}
%        75\\50 
%   \end{bmatrix}u,
% \end{align}
% which produces the discrete-time dynamical system in~\eqref{eq:mpcdiscrete} under a discretization time of $0.01$ seconds. %
Fig.\ref{fig:MPC}. depicts the level sets of the Lyapunov function. 
The Lyapunov function was found by all three refinement algorithms within one second. The Lyapunov-based refinement and the vector-field refinement, however, produce a greater number of cells than the naive refinement. %Also, the level set for all of these refinement approaches are almost the same. %
%Algorithm~\ref{alg:refinesearch} returns a valid Lyapunov ReLU neural network with $342$ neurons, in $82.34$ seconds. 
%The performance of all refinement methods in 2-D seems to be similar so far. In the next example, we will evaluate refinement methods using a 4-D example. 
\label{example:mpc}
\begin{figure}
    \centering
    \includegraphics[scale=0.48]{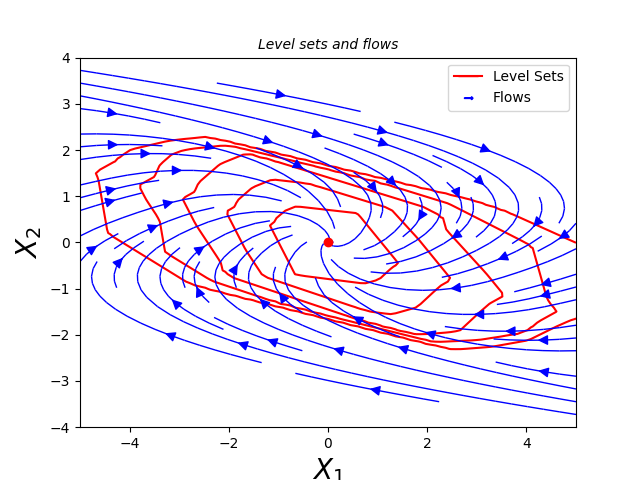}
    \caption{Selected level sets and flows for the Example \ref{example:mpc}.}
    \label{fig:MPC}
\end{figure}
\end{example}
\begin{example}[Inverted Pendulum\cite{farsi2022piecewise,chang2019neural,samanipour2023stability}]
\label{example:Pendulum}
It is common in the literature to use an inverted pendulum as an example with the following state-space model:
%An uncontrolled inverted pendulum is approximated by a single-hidden-layer ReLU neural network with 20 neurons. %
%The state-space dynamics of the pendulum are
\begin{align}
  \begin{bmatrix}
       \dot x_1 \\ \dot x_2
  \end{bmatrix} = \begin{bmatrix}
       x_2 \\ - \frac{c}{m} x_2 - g l^2  \sin(x_1)
  \end{bmatrix}+\begin{bmatrix}
       0 \\ \frac{1}{ml^2} 
  \end{bmatrix}u
\end{align}
where $m = 0.15$ kg, $l = 0.5$ m, c = $0.1 N s/rad$, and $g = 9.81$ m/s$^2$ \cite{zhou2022neural}.
%We are using $10000 $ samples to train the neural network using $\tt PyTorch 2.0.0$ with ADAM optimization. 
First, we used a single-hidden layer ReLU neural network consisting of 20 neurons in the region $\lVert x \rVert_{\infty} \leq 4$ to identify the uncontrolled dynamics. Subsequently, we designed a ReLU neural network controller as described in \cite{samanipour2023stability}.
By incorporating the ReLU NN controller into the system, we were able to achieve stability.
%This study examines the stability of the region associated with $\lVert x \rVert_{\infty} \leq 4$.
%The stability analysis was conducted on the same region, $\lVert x \rVert_{\infty} \leq 4$, as described in \cite{samanipour2023stability}. 
We searched for the $\PWA$ Lyapunov function using \textbf{Algorithm} \ref{alg:refinesearch} with the Vector-field refinement. 
%is used as a refinement technique in vertex-based method to find the Lyapunov function in this example, and
The results are compared with Linear-quadratic regulator (LQR)\cite{chang2019neural} and NN Lyapunov function\cite{zhou2022neural} in Fig.\ref{fig:invertedPend}. The Lyapunov function obtained using the proposed approach has a larger ROA. It is important to note that the valid region for \cite{chang2019neural} and \cite{zhou2022neural} is $\lVert x \rVert_{2} \leq 4$. 
\begin{figure}
    \centering
    \includegraphics[scale=0.48]{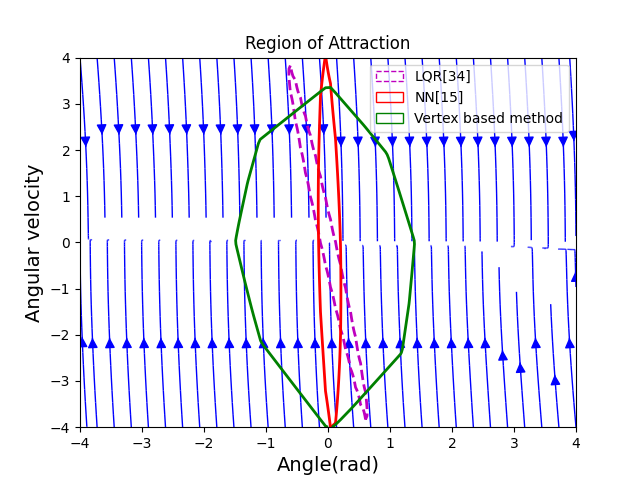}
    \caption{The ROA for the closed-loop inverted pendulum using LQR \cite{chang2019neural}, NN\cite{zhou2022neural} and vertex-based method.}
    \label{fig:invertedPend}
\end{figure}
\end{example}

Moreover, the computational time for each example is presented in the TABLE \ref{table:summary}. Having run each simulation ten times, the computational time is the average time elapsed. TABLE \ref{table:summary} provides the number of cells created by each refinement technique. The Vector field refinement performs better in terms of computation time and number of cells than the Lyapunov-based and naive approaches specifically in 4-D examples.
% \def\mtw{\textwidth}
% \begin{table}
% \vspace{1em}
% \centering
% \begin{tabular}{|p{0.12\mtw}|p{0.1\mtw}|p{0.08\mtw}|p{0.08\mtw}|}
% \cline{2-4}
% \multicolumn{1}{c}{}&\multicolumn{3}{|c|}{Average computational time(sec)} \\ \hline
% Example & Naive approach& Lyapunov based &  Vector field\\ \hline
% Example \ref{example:4D}[4-D] & timed-out & 1200 & 280 \\ \hline
% Example \ref{example:4D canonical} [4-D] & 1546 & 106 & 43\\\hline
% Example \ref{example:pathfollowing} [2-D] & 17.6 & 14.5 & 11.2 \\ \hline
% Example \ref{example:Multi-agent} [2-D]& 0.15 & 0.15 & 0.15\\\hline
% Example \ref{example:Flower_like dynamic} [2-D]& 1.8 & 1.8 & 1.86 \\ \hline
% Example \ref{example:mpc} [2-D] & 1.27 & 1.27 & 1.29 \\ \hline
% Example \ref{example:Pendulum} [2-D] & 23.3 & 17.4 & 16.1 \\ \hline
% \end{tabular}	
% \caption{Summary of examples of applying the proposed methods.}%; \mcr{S:} Strong stability}
% \label{table:summary}
% \end{table}
\begin{table}
\vspace{1em}
\centering
\begin{tabular}{|c|c|c|c|c|c|c|}
\hline 
Examples & \multicolumn{2}{c|}{Naive approach } & \multicolumn{2}{c|}{ Lyapunov based }& \multicolumn{2}{c|}{ Vector field } \\
\cline { 2 - 7 } & Time & cells & Time & Cells & Time & Cells \\
\hline Example \ref{example:4D}& Timed-out& 17764& 1200& 5874&280 &3086 \\
\hline Example \ref{example:4D canonical}&1546 &6943 &106 &2743 &43 &1054 \\
\hline Example \ref{example:pathfollowing}& 17.6& 705& 14.5 & 649&11.2 &  532 \\
\hline Example \ref{example:Multi-agent}& 0.15& 12& 0.15 & 12&0.15 &  12 \\
\hline Example \ref{example:Flower_like dynamic}& 1.8& 116& 1.8 & 116&1.86 &  120 \\
\hline Example \ref{example:mpc}& 1.27& 88 & 1.27 & 88 &1.29 &  96 \\
\hline Example \ref{example:Pendulum}& 23.3& 1107& 17.4 & 996 & 16.1& 956 \\

\hline
\end{tabular}
\caption{Summary of examples of applying the proposed methods. All times are in seconds.}% ; \mcr{S:} Strong stability}
\label{table:summary}
\end{table}
\section{Discussion}
\label{sec:limits}
\begin{figure}
    \centering
    \includegraphics[scale=0.5]{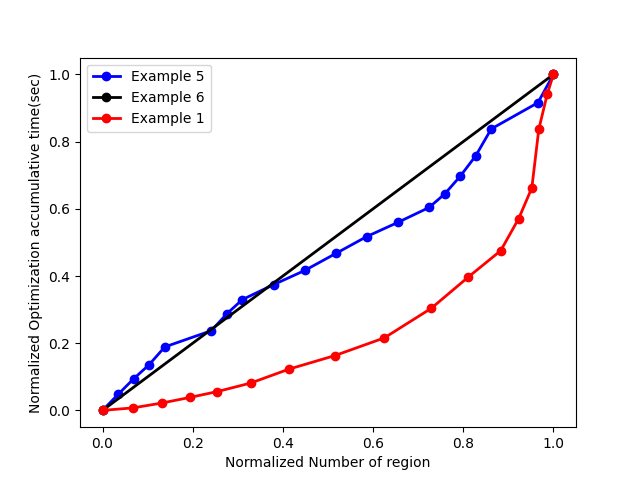}
    \caption{normalized Accumulative time for optimization vs. the normalized number of regions. The accumulative optimization time with respect to the number of cells for 2-D examples is almost linear; however, in the 4-D example, the accumulative time grows exponentially with respect to the number of cells.}
    \label{fig:opt}
\end{figure}

We have shown the effectiveness of our automated approach for stability verification through various examples. Our proposed refinement methods outperform existing techniques, and although our method does not specifically aim to maximize the region of attraction, our results are comparable to other methods. However, there are challenges to consider when applying this algorithm to a wider range of problems.
% We have demonstrated the effectiveness of our automated approach to stability verification through examples. We have demonstrated that the proposed refinement methods are superior to those available in the state-of-the-art. Additionally, our results demonstrated that, despite the fact that this method is not designed to maximize the region of attraction, its results are comparable with those obtained by other methods.  %
% The following sections discuss the challenges associated with making this algorithm applicable to a broader range of problems. 
\paragraph*{Limitations}
% To evaluate the computational complexity and performance of our proposed algorithm, it is important to consider the increase in the number of cells and optimization parameters during the refinement process. Generally, in a space $\mathbb{R}^n$, the number of cells $m$ should satisfy $m \geq 2^n$. In the simplest case, where the origin is surrounded by $2^n$ simplex cells, the optimization problem will have $2^n \times (n+1)$ parameters and $2^{n+1} \times n$ inequality constraints without considering equality constraints. However, if non-simplex cells are present in the partition, the number of constraints will increase accordingly. 
% %For instance, in a 6-D problem, the simplest case involves 448 parameters and 768 inequality constraints at the start of optimization, excluding equality constraints. 
% A large part of the optimization process depends on the $n$, which significantly influences the problem's computational complexity. As cells are further split, the problem becomes more complex, resulting in longer computational times for optimization. The algorithm may become bogged down in certain cases due to the increased complexity.
The computational complexity and performance of the proposed algorithm depend on the increase in the number of cells and optimization parameters during the refinement process. In a space $\mathbb{R}^n$, the number of cells should satisfy $m \geq 2^n$. The simplest case, where the origin is surrounded by $2^n$ simplex cells, results in an optimization problem with $2^n \times (n+1)$ parameters and $2^{n+1} \times n$ inequality constraints. The number of constraints increases with the presence of non-simplex cells. The computational complexity of the optimization process significantly depends on the dimensionality $n$, leading to longer computation times as cells are further divided. In some cases, the algorithm may encounter challenges and longer computation times due to increased complexity.
To compare the results of different examples in terms of computational time and the number of cells, we introduce the following concepts:
\begin{align}
T_{\text{opt}_i} &= \frac{\sum_{j=1}^i t_{\text{opt}j}}{\sum_{i=1}^N t_{\text{opt}_i}} \\
N{r_i} &= \frac{n_{r_i}}{\sum_{i=1}^N n_{r_i}}.
\end{align}
$T_ {opt}$ represents the normalized accumulative optimization time, $N_{r}$ indicates the normalized number of regions, $t_{opt}$ represents the time spent finding the solution with MOSEK, $n_{r}$ represents the number of regions, and $N$ represents the total number of iterations to solve the optimization problem. The subscripts $i$ and $j$ indicate the optimization iteration.
The relationship between the normalized accumulative optimization time ($T_{\text{opt}}$) and the normalized number of cells ($N_{r}$) is investigated in three different examples in Fig.\ref{fig:opt}. The graphs demonstrate almost linear behavior for Example \ref{example:mpc} and Example \ref{example:Flower_like dynamic}, while Example \ref{example:4D} exhibits an almost exponential trend. This indicates that increasing the number of cells could present a significant challenge for our proposed technique. Additionally, the refinement process may result in cells with nearly coplanar vertices, which can introduce numerical difficulties. It is essential to consider these complexities and challenges when applying our algorithm to various systems.
% Fig.\ref{fig:opt}. illustrates the relationship between the normalized accumulative optimization time $T_{\text{opt}}$ and the normalized number of cells $N_{r}$ for three different examples. In the case of Example \ref{example:Flower_like dynamic} and Example \ref{example:mpc}, the graph exhibits almost linear behavior. However, in Example \ref{example:4D}, the graph demonstrates an almost exponential trend. Therefore, increasing the number of cells could pose a significant challenge to our proposed technique. Furthermore, the refinement process may lead to cells with nearly coplanar vertices, which can introduce numerical difficulties.
% Considering the complexities and challenges associated with the refinement process is crucial when applying our algorithm to different systems. 
\section{Conclusion}
This paper presents a computational framework for obtaining valid $\PWA$ Lyapunov functions. The framework addresses the challenges of formulating the Lyapunov conditions as a linear optimization problem, which does not always guarantee a valid Lyapunov function. To overcome this limitation, two novel refinement methods are proposed, enhancing the flexibility of the candidate Lyapunov function. We used the Delaunay triangulation to automate the refinement process. We demonstrated that the proposed approach is effective based on experiments and comparisons with alternative approaches. The experiments successfully solve a 4-D example in a short time, highlighting the practicality and efficiency of the framework. The proposed framework offers a more effective method for generating valid $\PWA$ Lyapunov functions, offering flexibility through refinement methods and automating the process.
\bibliographystyle{IEEEtran}
\bibliography{main}
\end{document}